%% file: EXO-11-076_temp.tex
\pdfoutput=1

\documentclass[11pt,twoside,a4paper,cmspaper,final,collab]{cms-tdr}
\def\svnVersion{175880}\def\cmsCernNo{2012-203}\def\cmsCernDate{\today}\def\cmsMessage{Submitted to Physics Letters B}

\begin{document}\cmsNoteHeader{EXO-11-076}

\hyphenation{had-ron-i-za-tion}
\hyphenation{cal-or-i-me-ter}
\hyphenation{de-vices}

\RCS$Revision: 146355 $
\RCS$HeadURL: svn+ssh://svn.cern.ch/reps/tdr2/papers/EXO-11-076/trunk/EXO-11-076.tex $
\RCS$Id: EXO-11-076.tex 146355 2012-09-07 00:38:49Z alverson $
\newlength\cmsFigWidth
\ifthenelse{\boolean{cms@external}}{\setlength\cmsFigWidth{0.95\columnwidth}}{\setlength\cmsFigWidth{0.45\textwidth}}
\newlength\cmsFigWidthWide
\ifthenelse{\boolean{cms@external}}{\setlength\cmsFigWidthWide{0.95\columnwidth}}{\setlength\cmsFigWidthWide{0.8\textwidth}}
\cmsNoteHeader{EXO-11-076} 
\title{Search for heavy Majorana neutrinos in $\mu^\pm \mu^\pm$~+~jets and
$\Pe^\pm \Pe^\pm$~+~jets events in pp collisions at $\sqrt{s} = 7\TeV$}

\date{\today}

\abstract{
A search is performed for heavy Majorana neutrinos (N) using an
event signature defined by two same-sign charged leptons of the same flavour
and two jets. The data correspond to an integrated luminosity of 4.98\fbinv of pp collisions at a
centre-of-mass energy of 7\TeV collected with the CMS detector at the Large Hadron Collider. No excess of
events is observed beyond the expected standard model background and therefore upper limits are set on the square
of the mixing parameter, $|V_{\ell \mathrm{N}}|^2$, for $\ell = \Pe,\mu$, as a function of heavy Majorana-neutrino mass.
These are the first direct upper limits on the heavy Majorana-neutrino mixing for $m_\mathrm{N} > 90\GeV$.
}

\hypersetup{%
pdfauthor={CMS Collaboration},
pdftitle={Search for heavy  Majorana neutrinos in mu+mu+[mu-mu-] and e+e+[e-e-] events in pp collisions at sqrt(s) = 7 TeV},
pdfsubject={CMS},%
pdfkeywords={CMS, physics, heavy neutrino}}

\maketitle 

\section{Introduction}
The non-zero masses of neutrinos, confirmed from studies of their
oscillations among three species, provide the first evidence for physics beyond the standard model
(SM)~\cite{pdg}. The smallness of neutrino masses underscore the lack of a coherent
formulation for the generation of mass of elementary particles.
The leading theoretical candidate for accommodating neutrino masses is the so-called
``seesaw" mechanism \cite{seesaw1,seesaw2,seesaw3,seesaw4}, where
the smallness of the observed neutrino masses ($m_\nu$) is attributed to the largeness of a mass ($m_\mathrm{N}$)
of a new massive neutrino state N, with $m_\nu \approx y^2_\nu v^2/ m_\mathrm{N}$, where $y_\nu$ is a Yukawa coupling of $\nu$ to the Higgs field, and $v$ is the Higgs vacuum expectation value in the SM. In this model the SM neutrinos would also be Majorana particles. Owing to the new heavy
neutrino's Majorana nature, it is its own antiparticle, which allows processes that
violate lepton-number conservation by two units. Consequently, searches for heavy Majorana neutrinos
are of fundamental interest.

The phenomenology of searches for heavy Majorana neutrinos at hadron colliders has been considered by
many authors \cite{Maj_hadCol_1, Maj_hadCol_2, Maj_hadCol_3, Maj_hadCol_4, Maj_hadCol_5, TaoHan1,Aguila,TaoHan2}.
Our search follows the studies in Ref.~\cite{TaoHan1, Aguila} that
use a model-independent phenomenological approach, with $m_\mathrm{N}$ and $V_{\ell \mathrm{N}}$ as free parameters,
where $V_{\ell N}$ is a mixing parameter describing the mixing between the heavy Majorana neutrino and the SM neutrino
$\nu_\ell$ of flavour $\ell$. Previous direct searches for heavy Majorana neutrinos based on this model have been reported by the L3~\cite{l3} and DELPHI~\cite{delphi} Collaborations at the Large Electron-Positron Collider. They have searched for $\mathrm{Z} \to \nu_\ell \mathrm{N}$ decays and set limits
on $|V_{\ell \mathrm{N}}|^2$ as a function of $m_\mathrm{N}$ for heavy Majorana-neutrino masses up to approximately 90\GeV.
The ATLAS Collaboration at the Large Hadron Collider (LHC) has also reported limits on heavy Majorana neutrino production~\cite{ATLAS, ATLAS_2} in the  context of an effective Lagrangian approach~\cite{Wudka} and the Left-Right Symmetric Model~\cite{LRSM_1, LRSM_2}.
Indirect limits on $| V_{\Pe\mathrm{N}}|^2$ have been  obtained from the non-observation of neutrinoless double beta decay~\cite{beta_decay}, resulting in 90\% confidence level (CL) limits of
$| V_\mathrm{e N}|^2 / m_\mathrm{N}  < 7 \times  10^{-5}\TeV\mspace{0.5mu}^{-1}$. Precision electroweak measurements have been used to constrain the mixing parameters resulting in indirect 90\% CL limits of $| V_{\Pe\mathrm{N}}|^2 < 0.0066$,
$| V_{\mu \mathrm{N}}|^2 < 0.0060$, and $| V_{\tau \mathrm{N}}|^2 < 0.016$~\cite{EW_limits}.

\begin{figure}[htbp]
  \begin{center}
  \includegraphics[width=\cmsFigWidth]{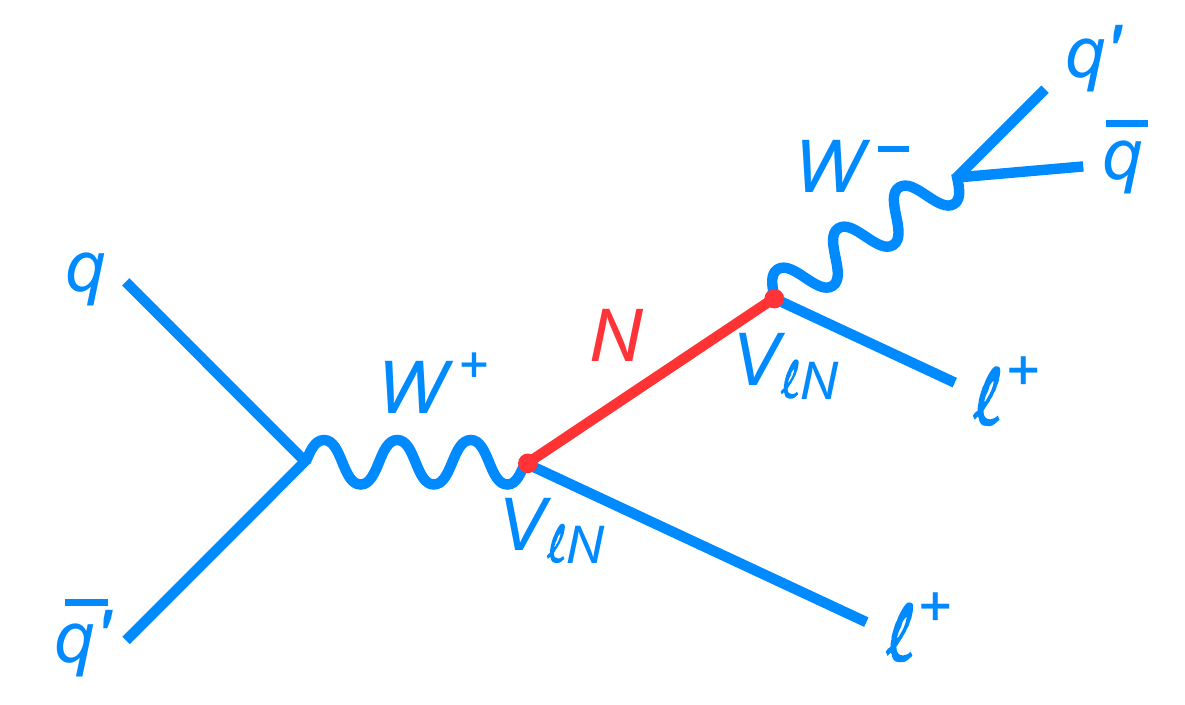}
  \caption{The lowest-order Feynman diagram for production of a heavy Majorana neutrino N.
  The charge-conjugate diagram also contributes and results in a $\ell^- \ell^- \mathrm{q \bar q^\prime}$ final state.
  }
  \label{fig:feynman}
  \end{center}
\end{figure}

We report on a search for the production of a heavy Majorana neutrino in proton-proton collisions at
a centre-of-mass energy of $\sqrt{s} = 7\TeV$ at the LHC using a set of data of integrated luminosity $4.98 \pm 0.11\fbinv$
collected with the Compact Muon Solenoid (CMS) detector.
The principal Feynman diagram for this process is shown in Fig.~\ref{fig:feynman}.
The heavy Majorana neutrino can decay to a lepton with positive or negative charge, leading to events containing two leptons with the same or opposite sign. Same-sign events have much lower backgrounds from SM processes and therefore provide an accessible signature of heavy Majorana neutrino production. We search for events  with two isolated leptons of same sign
and flavour ($\mu^\pm \mu^\pm$ or $\Pe^\pm \Pe^\pm$) and at least two accompanying jets.
Contributions from SM processes to such dilepton final states are very small and the background is dominated by processes such as multijet production, in which leptons from b-quark decays or from jets are misidentified as isolated prompt leptons.

\section{Signal simulation and data selection}
The production and decay process is simulated using the event generator described in
Ref.~\cite{Aguila} and implemented in \ALPGEN~\cite{ALPGEN}. We use the CTEQ5M parton distribution functions~\cite{CTEQ5}. Parton showering and hadronization are simulated using \PYTHIA~\cite{pythia}.  The Monte Carlo generated events are interfaced with CMS software, where \GEANTfour~\cite{Geant4} detector simulation, digitization of simulated electronic signals, and event reconstruction are performed. Monte Carlo simulated events are mixed with multiple minimum bias events with weights chosen using the distribution of the number of reconstructed primary vertices observed in data to ensure correct simulation of the number of interactions per bunch crossing (pileup). The average number of interactions per crossing in the data used in this analysis is approximately 9. The cross section for the process shown in Fig.~\ref{fig:feynman} for $| V_{\ell N}|^2 = 1$ has a value of 866\unit{pb} for $m_\mathrm{N} = 50\GeV$, which drops to 2.8\unit{pb} for $m_\mathrm{N} = 100\GeV$, and to 83\unit{fb} for $m_\mathrm{N} = 210\GeV$~\cite{Aguila}.

The CMS detector is described in detail in Ref.~\cite{CMS_detector}.
Its central feature is a superconducting solenoid, which provides a magnetic field of 3.8\unit{T} along
the direction of the counterclockwise rotating beam (as viewed from above the plane of the accelerator), taken as the $z$~axis of the detector coordinate system,
with the centre of the detector defined to be $z = 0$. The azimuthal angle $\phi$ is measured in the plane perpendicular to the $z$ axis, while the polar angle $\theta$ is measured with respect to this axis.
Muons are measured in four layers of gaseous ionization detectors embedded in the steel return yoke of the magnet, while all other particle detection systems are located inside the bore of the solenoid.
Charged particle trajectories are measured in a silicon pixel and strip tracker
covering $0 \le \phi \le 2 \pi$ in azimuth and $| \eta | < 2.5$,
where $\eta$ is the pseudorapidity, defined as $\eta = - \ln [\tan (\theta / 2)]$.
The tracker is surrounded by a lead tungstate
crystal electromagnetic calorimeter (ECAL) and a brass/scintillator hadron calorimeter
that are used to measure the energy of electrons, photons, and hadronic jets.
A two-level trigger system selects the most interesting events for analysis.

Dilepton triggers are used to select the signal sample. Depending on the average instantaneous luminosity of the LHC, dimuon events are recorded using a trigger requiring the presence of two muons, with transverse momenta (\pt) above 7\GeV for both muons in early data-taking runs, above 13\GeV for one muon and above 8\GeV for the second in later runs, or above 17\GeV for one muon and above 8\GeV for the second muon in most recent data.
Trigger efficiencies are measured using $\cPZ \to \mu^+ \mu^-$ and $\cPZ \to \Pep \Pem$ events selected in data, and are found to be $(96.0 \pm 2.0)$\% for muons and $(98.5 \pm 1.0)$\% for electrons.

Additional selections are performed offline to ensure the presence of well-identified muons, electrons, and jets. Events
are first required to have a well-reconstructed primary vertex based on charged tracks reconstructed in the tracking detectors.

Muon and electron candidates are required to have $| \eta | < 2.4$ and to be consistent with
originating from the primary interaction vertex. Muon candidates are reconstructed by matching tracks in
the silicon tracker to hits in the outer muon system, and are also required to satisfy specific track-quality and calorimeter-deposition requirements.
Electron candidates are reconstructed from energy depositions in the ECAL. These are matched to tracks in the silicon tracker and are required to satisfy shower distribution and cluster-track matching criteria. Electron candidates within $\Delta R = \sqrt{(\Delta \eta)^2 + (\Delta \phi)^2} < 0.4$ of a muon candidate are rejected to remove spurious electron candidates formed from the track of a muon that has an associated photon from bremsstrahlung. Electron candidates from photon conversions are suppressed by looking for a partner track and requiring that this track has no missing hits in the inner layers of the silicon tracker.

Electron and muon candidates must be isolated from other activity in the event by requiring their relative isolation ($I_\text{rel}$) to be less than 0.1. Here $I_\text{rel}$ is defined as the scalar sum of transverse track momenta and transverse calorimeter energy depositions present within $\Delta R < 0.3$ of the candidate's direction, excluding the candidate itself, divided by its transverse momentum.

Jets and the missing transverse energy in the event 
are reconstructed using the objects defined in the particle-flow method~\cite{pflow1, pflow2}. Jets are formed from clusters based on the anti-$k_\mathrm{T}$ algorithm~\cite{antiKT}, with a distance parameter of 0.5 and are required to
be within the pseudorapidity range $|\eta| < 2.5$ and  to have transverse momentum $\pt > 30$\GeV. At least two jets are required. The missing transverse energy is defined as the modulus of the negative of the vector sum of the transverse momenta of all reconstructed objects identified through the particle-flow algorithm.
The missing transverse energy is required to be less than 50\GeV.

Events in the muon channel are required to contain two same-sign muons, one  with $\pt$ greater than 20\GeV and the other with $\pt$ greater than 10\GeV. Events with an opposite-sign third muon that combines with one of the other candidate muons to give
a $\mu^+ \mu^-$ invariant mass within the window for a $\cPZ$ boson of 76--106\GeV are excluded.
In the electron channel, events are required to contain two same-sign electrons, one  with $\pt$ greater than 20\GeV and one with $\pt$ greater than 10\GeV. Events containing any third electron candidate are rejected.
Overall signal acceptance includes trigger efficiency, geometrical acceptance, and all selection criteria.
In the muon channel, the overall acceptance for heavy Majorana neutrino events ranges between $0.43\%$ for $m_\mathrm{N} = 50$\GeV to $29\%$ for $m_\mathrm{N} = 210$\GeV. For the electron channel, the corresponding efficiency changes from 0.40\% to 21\% for these masses. The lower acceptance at low $m_\mathrm{N}$ is due to the smaller average $\pt$ of the jets and leptons in these events.

\section{Background estimation}
There are three potential sources of same-sign dilepton backgrounds.  The first and most important originates from events containing leptons from b-quark decays or generic jets that are misidentified as leptons.
Examples of this background include: (i) multijet production in which two jets are misidentified as leptons;
(ii) $\PW(\to \ell \nu)$~+~jets events in which one of the jets is misidentified as a lepton; and
(iii) $\ttbar$ decays in which one of the top quarks decays giving a prompt isolated lepton
($\cPqt \to \PW \cPqb \to \ell \nu_\ell \cPqb$), and the other lepton of same charge arises from a b-quark decay. From Monte Carlo studies we find that the dominant contribution to this background is from multijet production, with the sum of $\PW(\to \ell \nu)$~+~jets and $\ttbar$ events comprising approximately 15--35\% of the total misidentified lepton background. These backgrounds are estimated using control samples in collision data as described below.

To estimate the misidentified lepton background, an independent data sample enriched in multijet events is used to
calculate the probability for a jet that passes minimal lepton selection requirements to also pass the more stringent requirements used to define selected leptons.
The lepton candidates  passing the less stringent requirements are referred to as ``loose leptons'' and their  misidentification probability  is calculated as a function of transverse momentum and pseudorapidity.
This probability is used as a weight in the calculation of the background in
events that pass all the signal selections except that one or both leptons
fail the tight criteria (used to select the leptons in signal events). This sample is referred to as the ``orthogonal'' sample.

The misidentification probability is applied to the orthogonal sample by counting
the number of events in which one lepton passes the tight criteria, while the other lepton fails the
tight selection but passes the loose selection ($N_{n\overline{n}}$), and the number of events in which both
leptons fail the tight selection, but pass the loose criteria ($N_{\overline{n}\overline{n}}$). The total contribution to the signal sample (i.e. the number of events when both leptons pass the tight selection, $N_{nn}$), is then obtained by weighting events of type $n \overline{n}$ and $\overline{n} \overline{n}$ by the appropriate misidentification probability factors. To account for double counting we correct for $\overline{n} \overline{n}$ events that can also be $n \overline{n}$.

In the muon channel, loose muons are defined by relaxing the muon isolation requirement from $I_\text{rel} < 0.1$ (used to select signal events) to $I_\mathrm{rel} <0.8$. In the electron channel, loose electrons are defined by relaxing the isolation from 0.1 to 0.6, and by removing a requirement on transverse impact parameter normally used for tight electrons.

We evaluate the method used to estimate the background from misidentified leptons by checking the procedure using Monte Carlo simulated event samples in which the true background is known. The misidentification probabilities are obtained from multijet events and are used to estimate the misidentified lepton backgrounds in $\ttbar$, \PW~+~jets, and multijet events by applying the background estimation method described above.
The differences between the estimated backgrounds and the true number of events in the Monte Carlo samples is used as input to the overall systematic uncertainty.

The overall systematic uncertainty on the misidentified lepton background is determined from
the variation of the background estimate with the loose lepton definition and the variation with the leading jet $\pt$
requirement in the data samples used to measure the misidentification probability, as well as from the independent Monte Carlo validation studies described above. The overall uncertainty is found to be 35\%.

The second contribution to the background is from $\ell^+ \ell^-$ events in which the charge of one of the leptons is mismeasured. Since the charge mismeasurement probability is very small for muons, this background is significant only in the electron channel. 
The background from mismeasurement of electron charge is estimated through probabilities calculated in Monte Carlo studies
of $\Pep \Pem$ events that pass all selections for signal, except the same-sign requirement.
The average electron mismeasurement probability is found to be $(3.3 \pm 0.2) \times 10^{-4}$  in the ECAL barrel region ($|\eta| < 1.5$) and $(2.9  \pm 0.1) \times 10^{-3}$ in the ECAL endcap region ($1.5 < |\eta| < 2.5$). To validate the charge mismeasurement probability, we select a control sample of $\cPZ \to \Pep \Pem$ events in data, requiring an  $\Pep \Pem$ invariant mass between 76 and 106\GeV.
We use the difference between the predicted and the observed number of $\Pe^\pm \Pe^\pm$ events, including uncertainties, to set a systematic uncertainty of 25\% on the background from charge mismeasurement.

The third background source is the irreducible background from SM production of two genuine isolated leptons of the same sign, which can originate from sources such as $\cPZ\cPZ$, $\PW\cPZ$, and $\PW\gamma$ diboson production, $\cPqt \cPaqt \PW$, double \PW-strahlung $\PW^\pm \PW^\pm \cPq \cPq$ processes, and double-parton scattering (two $\cPq \cPq^\prime \to \PW)$. These have relatively small cross sections, and are consequently estimated using Monte Carlo simulations. We use \PYTHIA\ to simulate $\cPZ\cPZ$ and $\PW\cPZ$ production and \MADGRAPH~\cite{madgraph} for the remaining processes.

\section{Systematic uncertainties}

The sources of systematic uncertainty associated with signal efficiency and background estimates can be summarized
as follows.
\begin{itemize}
\item[(1)] The systematic uncertainty on the integrated luminosity is 2.2\%~\cite{lumi}.
\item[(2)] The systematic uncertainty from choice of parton distribution functions is estimated from Monte Carlo simulations following the PDF4LHC recommendations~\cite{pdf} and is found to be 6\% of the signal yield.
\item[(3)] The hard-scattering scale in the \ALPGEN Monte Carlo generator is varied from the nominal value of $Q^2$
to $4Q^2$ and $Q^2/4$. The resulting uncertainty on the signal yield is 1\%.
\item[(4)] The jet energy scale is changed by its estimated uncertainty~\cite{jes} resulting in a systematic uncertainty of between $3.3\%$ for high heavy Majorana-neutrino mass ($m_\mathrm{N} = 210\GeV$) and $14.2\%$ at low mass
($m_\mathrm{N} = 50\GeV$). For low mass events the jets from the heavy Majorana neutrino decay have lower average \pt, leading to larger uncertainties.
\item[(5)] The systematic uncertainty due to the uncertainty on the jet energy resolution~\cite{jes} is determined
to be  0.2\%--1\%, depending on $m_\mathrm{N}$.
\item[(6)] The uncertainty in modeling event pileup is studied in the Monte Carlo simulations and found to be $1\%$.
\item[(7)] The systematic  uncertainty on the estimate of misidentified lepton background is 35\%.
\item[(8)] The systematic uncertainty on the background from mismeasurement of electron charge is 25\%.
\item[(9)] The systematic uncertainties on normalizations of irreducible SM backgrounds are: 6\% for $\PW\cPZ$ and $\cPZ\cPZ$~\cite{diboson_xs}; 10\% for $\PW \gamma$~\cite{diboson_xs}; and 50\% for the other processes, determined by varying the $Q^2$ scale and parton distribution functions in Monte Carlo simulations.
\end{itemize}

In the muon channel the systematic uncertainty due to the muon trigger, as indicated above, is based on studies of $\cPZ \to \Pgmp \Pgmm$ events in collision data, and is determined to be 2\% per muon.  The offline muon selection efficiency is taken from Monte Carlo simulation, and cross-checked with data using studies of $\cPZ \to \Pgmp\Pgmm$ events. The  efficiencies measured in data and Monte Carlo simulation are found to be in agreement within uncertainties; the systematic uncertainty associated with the small differences is 2\%. The overall systematic uncertainty due to the muon trigger and selection is 4\%. In the electron channel the systematic uncertainty from the trigger and electron selections is determined in a similar way, and found to be 10\%.

\section{Results and discussion}

After applying all the final selections we observe 65 events in data in the muon channel and expect a total SM background of
$70 \pm 4~\stat \pm 22~\syst$ events, with the dominant contribution of
$63 \pm 4~\stat \pm 22~\syst$ events arising from the misidentified muon background. The data are in agreement with the estimated background.
In the electron channel we observe 201 events in data, and estimate the total SM  background as
$219 \pm 6~\stat \pm 62~\syst$ events, with the dominant contribution of $177 \pm 5~\stat \pm 62~\syst$ events
arising from the misidentified electron background. The data are again in agreement with the estimated background.
The final estimates for the two channels are given in Table~\ref{table:yields}.
\begin{table*}[t!bph]
  \topcaption{Observed event yields and estimated backgrounds in the muon and
  electron channel.
  Also shown are the expected number of signal events for two heavy Majorana-neutrino masses of
  130 and 210\GeVcc, for a mixing parameter $\left| V_{\ell N} \right|^2 = 0.1$. The sets of first and
  second uncertainties on the background and signal estimates correspond, respectively, to statistical and
  systematic contributions.
  For the irreducible SM backgrounds and the expected signals, the first uncertainty is due to
  the statistical error associated with the finite size of the Monte Carlo event samples used.}
  \label{table:yields}
\begin{center}
\begin{tabular}{l l l l}
\hline \hline
Source & $\mu^\pm \mu^\pm$ & $\Pe^\pm \Pe^\pm$ \\
\hline
Irreducible SM backgrounds:  & \\
~~~~~WZ               & $3.2   \pm 0.3  \pm 0.2$    & $4.9  \pm 0.3  \pm 0.3$ \\
~~~~~ZZ               & $1.0   \pm 0.1  \pm 0.1$    & $2.1  \pm 0.1  \pm 0.1$ \\
~~~~~W$\gamma$        & $0.75  \pm 0.27 \pm 0.07$   & $1.7  \pm 0.4  \pm 0.2$ \\
~~~~~$\ttbar \PW$
                   & $1.06  \pm 0.05 \pm 0.53$    & $0.62 \pm 0.04 \pm 0.31$ \\
~~~~~W$^+$W$^+$qq                   & $0.76  \pm 0.06 \pm 0.38$   & $0.73 \pm 0.07 \pm 0.37$ \\
~~~~~W$^-$W$^-$qq                   & $0.45  \pm 0.03 \pm 0.23$   & $0.27 \pm 0.02 \pm 0.13$ \\
~~~~~Double-parton ~W$^\pm$W$^\pm$  & $0.07  \pm 0.02 \pm 0.04$   & $0.19 \pm 0.03 \pm 0.10$ \\
Total irreducible SM background  & $7.3 \pm 0.4 \pm 0.7$   & $10.6 \pm 0.6 \pm 0.6$ \\
Charge mismeasurement background & $0^{+0.2}_{-0}$ 	  & $31.9 \pm  2.7 \pm 8.0$	\\
Misidentified lepton background & $63.1 \pm 4.2 \pm 22.1$  & $176.8 \pm 4.7 \pm 61.9$ \\
\hline
Total background     & $70 \pm 4 \pm 22$     & $219 \pm 6 \pm 62$ \\
Data                 & $65$  & $201$ \\
\hline
Expected signal: & & \\
~~~~~$m_N = 130$\GeVcc, $\left| V_{\ell N} \right|^2 = 0.1$	& $58 \pm 1 \pm 4$				&	$39 \pm 1 \pm 3$ \\
~~~~~$m_N = 210$\GeVcc, $\left| V_{\ell N} \right|^2 = 0.1$	&	$12.0 \pm 0.1 \pm 0.8$	& $8.5 \pm 0.1 \pm 0.6$ \\
\hline
\end{tabular}
\end{center}
\end{table*}

\begin{figure}[tbh]
\begin{center}
	\includegraphics[width=\cmsFigWidth]{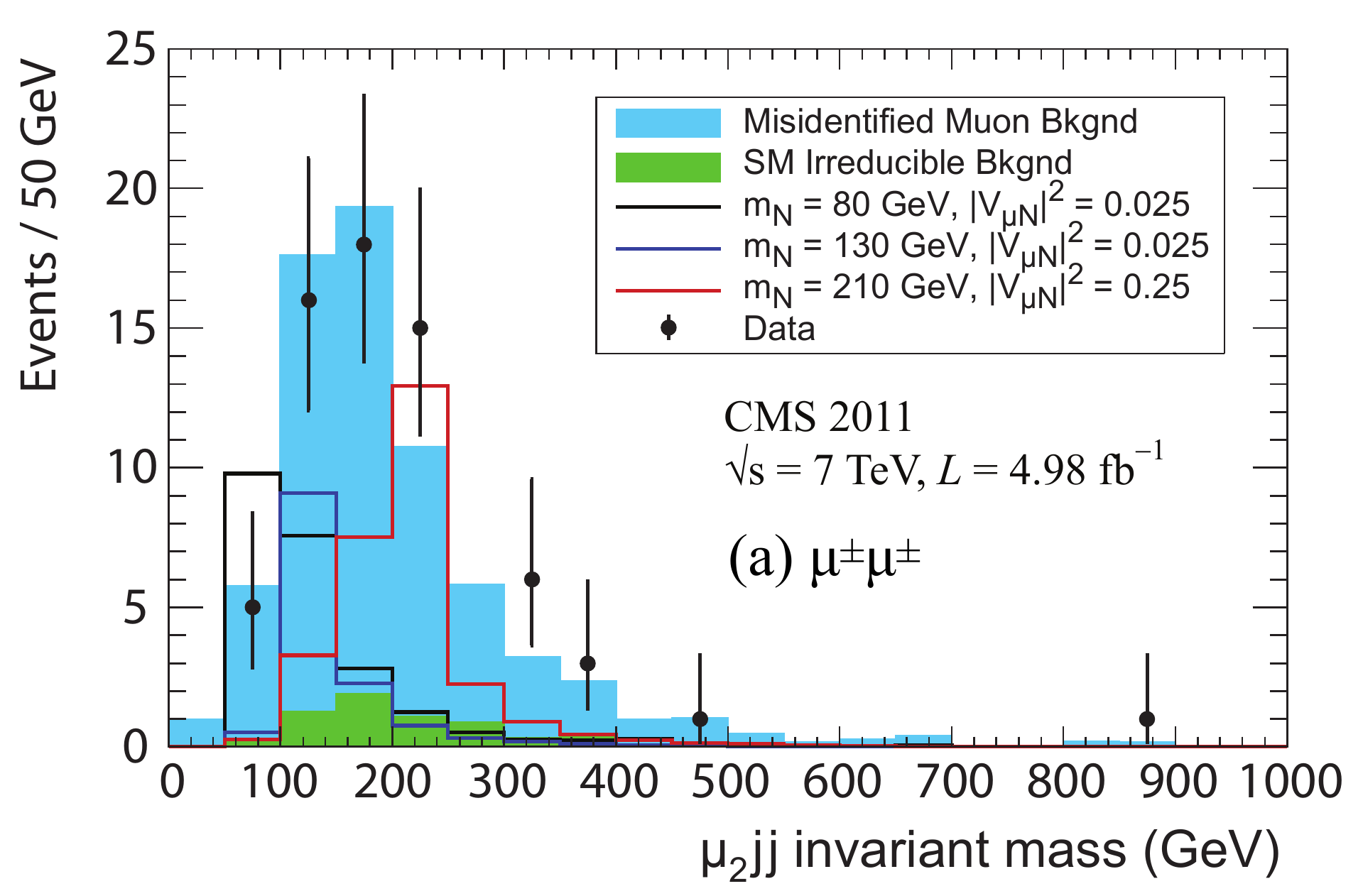}
	\includegraphics[width=\cmsFigWidth]{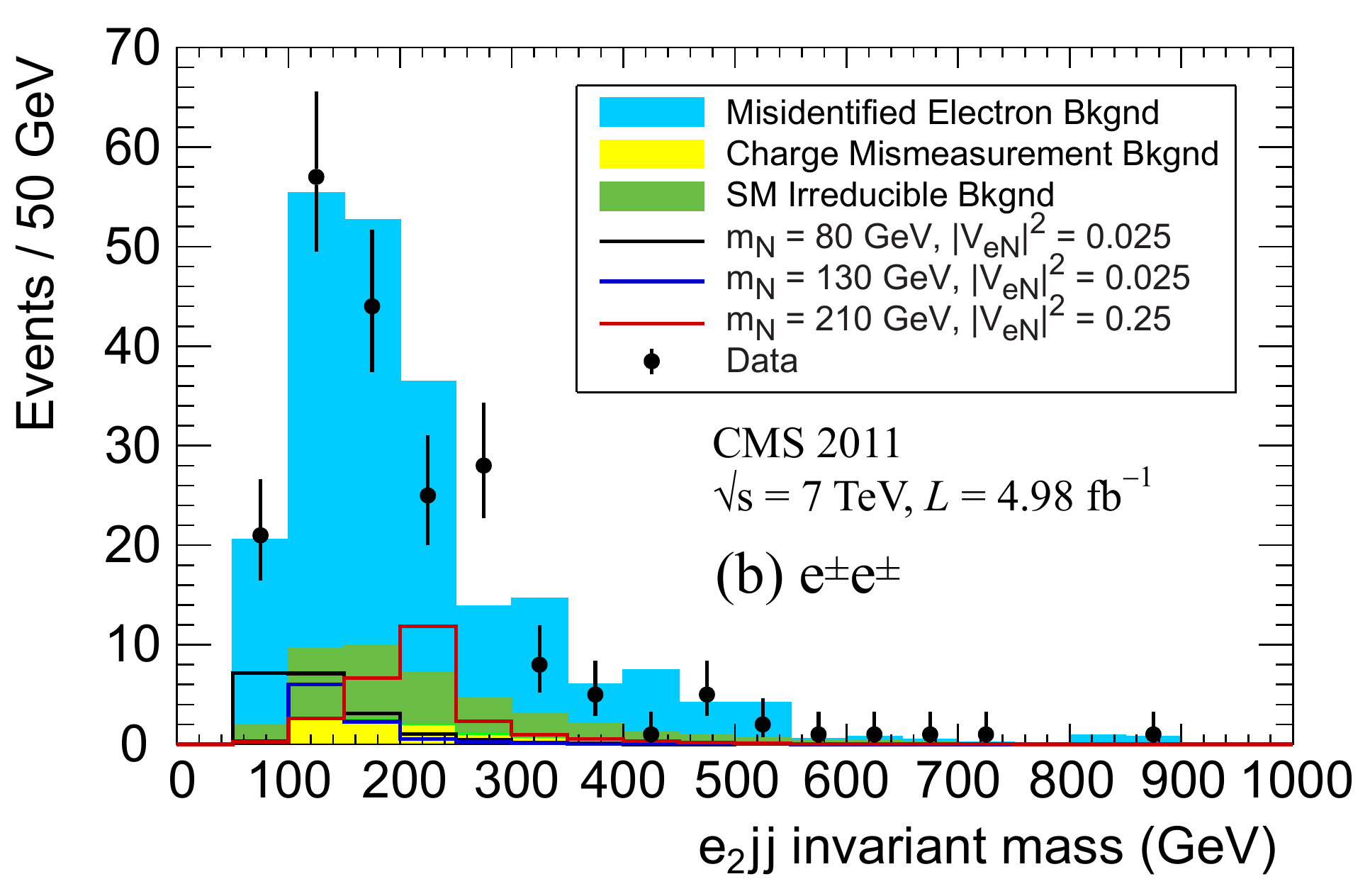}
	\caption{Invariant mass of the second leading $\pt$ lepton and the two leading jets for events passing the
	signal selection. The plots show the data, standard model backgrounds, and three choices for the
	heavy Majorana-neutrino signal:  $m_N = 80$\GeVcc, $\left| V_{\ell N} \right|^2 = 0.025$,
	$m_N = 130$\GeVcc, $\left| V_{\ell N} \right|^2 = 0.025$, and
	$m_N = 210$\GeVcc, $\left| V_{\ell N} \right|^2 = 0.25$.
	(a) Distributions for $\mu^\pm \mu^\pm$ events; (b) distributions for $\Pe^\pm \Pe^\pm$ events.}
	\label{fig:mass}
\end{center}
\end{figure}

Figure~\ref{fig:mass} shows the distributions of invariant mass of the second highest $\pt$ lepton and the two leading jets in the event for data, standard model backgrounds, and three choices for the Majorana neutrino signal:
$m_N = 80$\GeVcc, $\left| V_{\ell N} \right|^2 = 0.025$, $m_N = 130$\GeVcc, $\left| V_{\ell N} \right|^2 = 0.025$,
and $m_N = 210$\GeVcc, $\left| V_{\ell N} \right|^2 = 0.25$. From Monte Carlo studies, the choice of second highest $\pt$ lepton is found to give  better performance for reconstruction of the Majorana neutrino mass compared with taking either the leading $\pt$ lepton or randomly picking one of the leptons.

\begin{figure}[hbtp]\begin{center}
  \includegraphics[width=\cmsFigWidth]{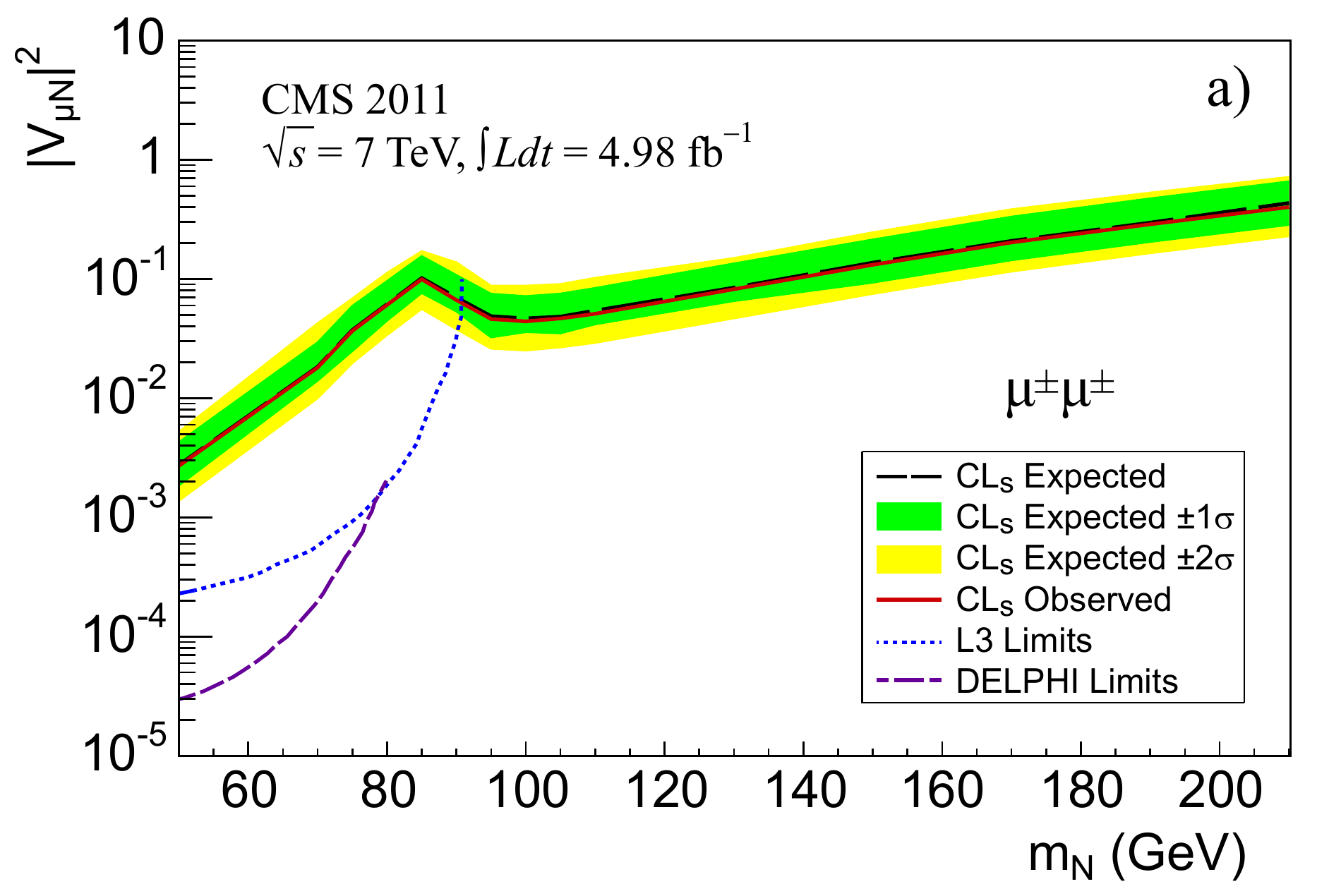}
  \includegraphics[width=\cmsFigWidth]{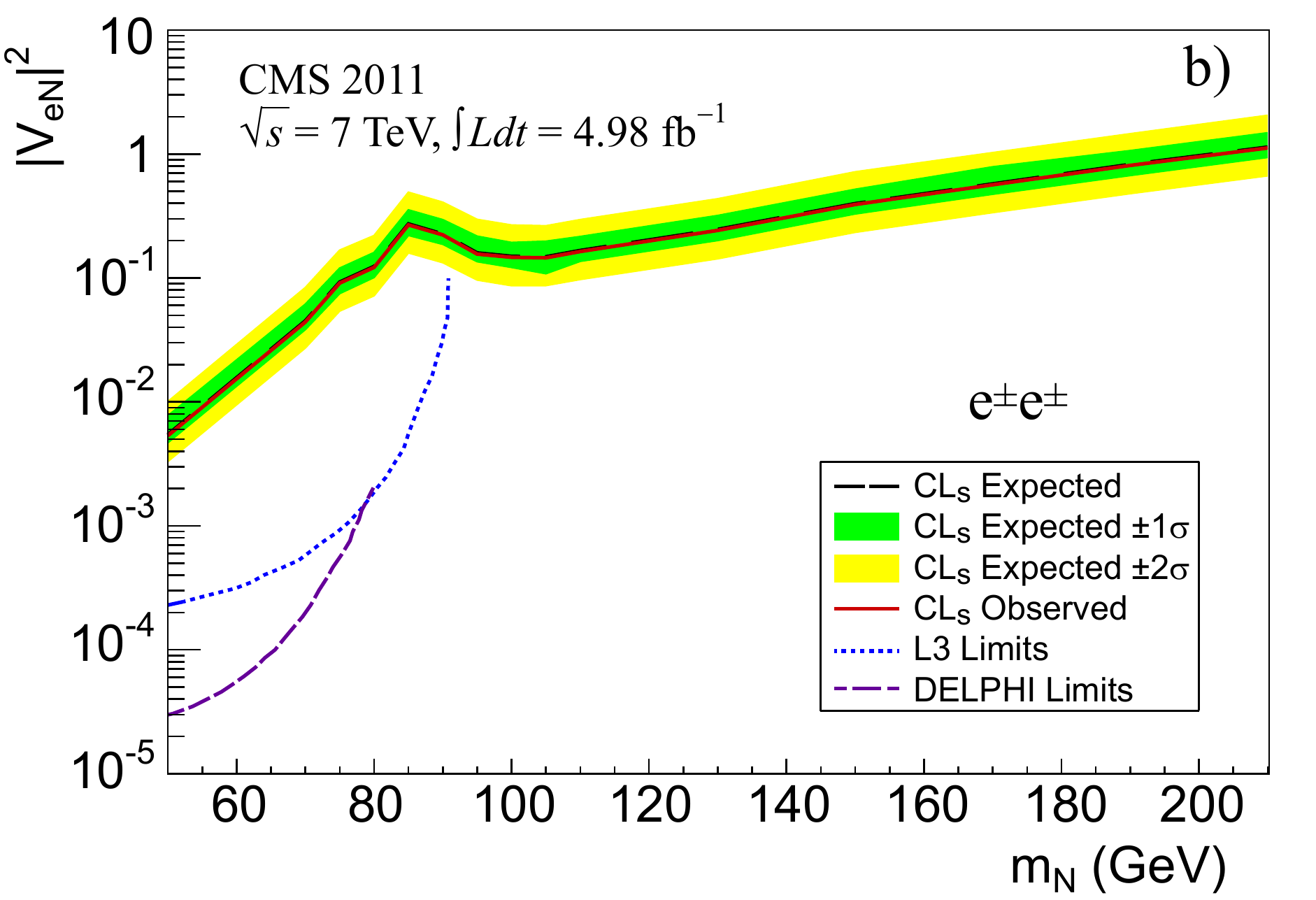}
  \caption{Exclusion region at 95\% CL in the square of the heavy Majorana-neutrino mixing parameter
  as a function of the heavy Majorana-neutrino mass: (a) $| V_{\mu \mathrm{N}} |^2$ vs $m_\mathrm{N}$;
  (b) $| V_{\Pe \mathrm{N}} |^2$ vs. $m_\mathrm{N}$. The long-dashed black line is the expected upper limit, with one and two
  standard-deviation bands shown in dark green and light yellow, respectively. The solid red line is the
  observed upper limit, and is very close to the expected limit such that the two curves almost overlap.
  Also shown are the upper limits from L3~\cite{l3} and DELPHI~\cite{delphi}. The regions above the exclusion
  lines are ruled out at 95\% CL.}
\label{fig:excl}
\end{center}\end{figure}

We see no evidence for a significant excess in the data beyond the backgrounds predicted from the SM
and set 95\% CL exclusion limits  on the square of the heavy Majorana-neutrino
mixing parameter as a function of $m_\mathrm{N}$, using the CL$_s$ method~\cite{CLs1, CLs2, CLs3} based on the event yields shown in Table~~\ref{table:yields}. In the
muon channel analysis we set limits on $| V_{\mu \mathrm{N}} |^2$ as a function of $m_\mathrm{N}$, under the assumption
$| V_{\Pe \mathrm{N}} |^2 = | V_{\tau \mathrm{N}} |^2 = 0$. In the electron channel analysis we set limits on
$| V_{\Pe \mathrm{N}} |^2$ as a function of $m_\mathrm{N}$ under the assumption
$| V_{\mu \mathrm{N}} |^2 = | V_{\tau \mathrm{N}} |^2 = 0$.
Figure~\ref{fig:excl}(a) shows the resulting upper limits in the  muon channel
($| V_{\mu \mathrm{N}} |^2$~vs.~$m_\mathrm{N}$), while Fig.~\ref{fig:excl}(b) shows the upper limits in the electron channel ($| V_{\Pe \mathrm{N}} |^2$~vs.~$m_\mathrm{N}$). These are the first direct upper limits on the heavy Majorana-neutrino mixing for $m_\mathrm{N} > 90$\GeV.

For low $m_\mathrm{N}$ the limits in both channels are less stringent than the existing limits from DELPHI and L3 shown in Figs.~\ref{fig:excl}(a) and ~\ref{fig:excl}(b), due to the higher backgrounds at the LHC. However, the DELPHI and L3 limits are derived from $\cPZ \to \nu_\ell N$ and are restricted to masses below approximately 90\GeV. The limits reported here extend well beyond this mass. For $m_\mathrm{N} = 90$\GeV we find $| V_{\mu \mathrm{N}} |^2 < 0.07$ and
$| V_{\Pe \mathrm{N}} |^2 < 0.22$. At $m_\mathrm{N} = 210$\GeV  we find $| V_{\mu \mathrm{N}} |^2 < 0.43$, while for
$| V_{\Pe \mathrm{N}} |^2$ the limit reaches 1.0 at a mass of 203\GeV.

\section{Summary}
A search for heavy Majorana neutrinos in
$\mu^\pm \mu^\pm$  and $\Pe^\pm \Pe^\pm$ events has been performed using a set of data corresponding to
5.0\fbinv of pp collisions at a centre-of-mass energy of 7\TeV. No excess of events beyond the
standard model background prediction is found. Upper limits at the 95\% CL are set on the
square of the heavy Majorana-neutrino mixing parameter, $| V_{\ell \mathrm{N}} |^2$, for $\ell = \Pe,\mu$, as a function of heavy Majorana-neutrino mass, as shown in Figs.~\ref{fig:excl}(a) and \ref{fig:excl}(b).
For $m_\mathrm{N} = 90$\GeV the limits are $| V_{\mu \mathrm{N}} |^2 < 0.07$ and
$| V_{\Pe \mathrm{N}} |^2 < 0.22$. At $m_\mathrm{N} = 210$\GeV  the limits are $| V_{\mu \mathrm{N}} |^2 < 0.43$, while for $| V_\mathrm{e N} |^2$ the limit reaches 1.0 at a mass of 203\GeV.
These are the first direct upper limits on the heavy Majorana-neutrino mixing for $m_\mathrm{N} > 90$\GeV.

\section*{Acknowledgements}

We congratulate our colleagues in the CERN accelerator departments for the excellent performance of the LHC machine. We thank the technical and administrative staffs at CERN and other CMS institutes, and acknowledge support from: FMSR (Austria); FNRS and FWO (Belgium); CNPq, CAPES, FAPERJ, and FAPESP (Brazil); MES (Bulgaria); CERN; CAS, MoST, and NSFC (China); COLCIENCIAS (Colombia); MSES (Croatia); RPF (Cyprus); MoER, SF0690030s09 and ERDF (Estonia); Academy of Finland, MEC, and HIP (Finland); CEA and CNRS/IN2P3 (France); BMBF, DFG, and HGF (Germany); GSRT (Greece); OTKA and NKTH (Hungary); DAE and DST (India); IPM (Iran); SFI (Ireland); INFN (Italy); NRF and WCU (Korea); LAS (Lithuania); CINVESTAV, CONACYT, SEP, and UASLP-FAI (Mexico); MSI (New Zealand); PAEC (Pakistan); MSHE and NSC (Poland); FCT (Portugal); JINR (Armenia, Belarus, Georgia, Ukraine, Uzbekistan); MON, RosAtom, RAS and RFBR (Russia); MSTD (Serbia); SEIDI and CPAN (Spain); Swiss Funding Agencies (Switzerland); NSC (Taipei); TUBITAK and TAEK (Turkey); STFC (United Kingdom); DOE and NSF (USA).

\bibliography{auto_generated}   

\cleardoublepage \appendix\section{The CMS Collaboration \label{app:collab}}\begin{sloppypar}\hyphenpenalty=5000\widowpenalty=500\clubpenalty=5000\input{EXO-11-076-authorlist.tex}\end{sloppypar}
\end{document}

%% file: EXO-11-076-authorlist.tex
\textbf{Yerevan Physics Institute,  Yerevan,  Armenia}\\*[0pt]
S.~Chatrchyan, V.~Khachatryan, A.M.~Sirunyan, A.~Tumasyan
\vskip\cmsinstskip
\textbf{Institut f\"{u}r Hochenergiephysik der OeAW,  Wien,  Austria}\\*[0pt]
W.~Adam, E.~Aguilo, T.~Bergauer, M.~Dragicevic, J.~Er\"{o}, C.~Fabjan\cmsAuthorMark{1}, M.~Friedl, R.~Fr\"{u}hwirth\cmsAuthorMark{1}, V.M.~Ghete, J.~Hammer, N.~H\"{o}rmann, J.~Hrubec, M.~Jeitler\cmsAuthorMark{1}, W.~Kiesenhofer, V.~Kn\"{u}nz, M.~Krammer\cmsAuthorMark{1}, I.~Kr\"{a}tschmer, D.~Liko, I.~Mikulec, M.~Pernicka$^{\textrm{\dag}}$, B.~Rahbaran, C.~Rohringer, H.~Rohringer, R.~Sch\"{o}fbeck, J.~Strauss, A.~Taurok, W.~Waltenberger, G.~Walzel, E.~Widl, C.-E.~Wulz\cmsAuthorMark{1}
% ...
\vskip\cmsinstskip
\dag:~Deceased\\
1:~~Also at Vienna University of Technology, Vienna, Austria\\
% ...


%% file: EXO-11-076_temp.bbl
\providecommand{\href}[2]{#2}\begingroup\raggedright\begin{thebibliography}{10}%
\makeatletter
\providecommand{\hrefCMSnoop }[0]{\@secondoftwo}%
\makeatother
\providecommand{\doi}{\texttt{doi:}\begingroup \urlstyle{tt}\Url}

\bibitem{pdg}
\hrefCMSnoop {} {J.~Beringer {et~al.}, ``{Review of particle physics}'',}
  \textit{ Phys. Rev. D} \textbf{ 86} (2012) 010001,
  \href{http://dx.doi.org/10.1103/PhysRevD.86.010001}{\doi{10.1103/PhysRevD.86.010001}}.
  See the review section on Neutrino Mass, Mixing, and Oscillations, and
  references therein.

\bibitem{seesaw1}
\hrefCMSnoop {} {P.~Minkowski, ``{$\mu \to e \gamma$ at a Rate of One Out of
  1-Billion Muon Decays?}'',} \textit{ Phys. Lett. B} \textbf{ 67} (1977) 421,
\href{http://dx.doi.org/10.1016/0370-2693(77)90435-X}{\doi{10.1016/0370-2693(77)90435-X}}.

\bibitem{seesaw2}
\hrefCMSnoop {} {M.~Gell-Mann, P.~Ramond, and R.~Slansky, ``Complex Spinors and
  Unified Theories'',} in \textit{ Supergravity: proceedings of the
  Supergravity Workshop at Stony Brook}, P.~V. Nieuwenhuizen and D.~Z.
  Freedman, eds., p.~315.
\newblock North-Holland, 1979.

\bibitem{seesaw3}
\hrefCMSnoop {} {T.~Yanagida} in \textit{ Proceedings of the Workshop on the
  Unified Theory and the Baryon Number in the Universe}, O.~Sawada and
  A.~Sugamoto, eds., p.~95.
\newblock National Laboratory for High Energy Physics (KEK), 1979.

\bibitem{seesaw4}
\hrefCMSnoop {} {R.~N. Mohapatra and G.~Senjanovic, ``{Neutrino Mass and
  Spontaneous Parity Violation}'',} \textit{ Phys. Rev. Lett.} \textbf{ 44}
  (1980) 912,
\href{http://dx.doi.org/10.1103/PhysRevLett.44.912}{\doi{10.1103/PhysRevLett.44.912}}.

\bibitem{Maj_hadCol_1}
\hrefCMSnoop {} {W.-Y. Keung and G.~Senjanovic, ``{Majorana Neutrinos And The
  Production Of The Right-Handed Charged Gauge Boson}'',} \textit{ Phys. Rev.
  Lett.} \textbf{ 50} (1983) 1427,
\href{http://dx.doi.org/10.1103/PhysRevLett.50.1427}{\doi{10.1103/PhysRevLett.50.1427}}.

\bibitem{Maj_hadCol_2}
\hrefCMSnoop {} {D.~A. Dicus, D.~D. Karatas, and P.~Roy, ``{Lepton
  nonconservation at supercollider energies}'',} \textit{ Phys. Rev. D}
  \textbf{ 44} (1991) 2033,
\href{http://dx.doi.org/10.1103/PhysRevD.44.2033}{\doi{10.1103/PhysRevD.44.2033}}.

\bibitem{Maj_hadCol_3}
\hrefCMSnoop {} {A.~Datta, M.~Guchait, and A.~Pilaftsis, ``{Probing lepton
  number violation via majorana neutrinos at hadron supercolliders}'',}
  \textit{ Phys. Rev. D} \textbf{ 50} (1994) 3195,
  \href{http://dx.doi.org/10.1103/PhysRevD.50.3195}{\doi{10.1103/PhysRevD.50.3195}},
\href{http://www.arXiv.org/abs/hep-ph/9311257}{\texttt{ arXiv:hep-ph/9311257}}.

\bibitem{Maj_hadCol_4}
F.~M.~L. Almeida~Jr.\hrefCMSnoop {} { {et~al.}, ``{On a signature for heavy
  Majorana neutrinos in hadronic collisions}'',} \textit{ Phys. Rev. D}
  \textbf{ 62} (2000) 075004,
  \href{http://dx.doi.org/10.1103/PhysRevD.62.075004}{\doi{10.1103/PhysRevD.62.075004}},
\href{http://www.arXiv.org/abs/hep-ph/0002024}{\texttt{ arXiv:hep-ph/0002024}}.

\bibitem{Maj_hadCol_5}
O.~Panella\hrefCMSnoop {} { {et~al.}, ``{Signals of heavy Majorana neutrinos at
  hadron colliders}'',} \textit{ Phys. Rev. D} \textbf{ 65} (2002) 035005,
  \href{http://dx.doi.org/10.1103/PhysRevD.65.035005}{\doi{10.1103/PhysRevD.65.035005}},
\href{http://www.arXiv.org/abs/hep-ph/0107308}{\texttt{ arXiv:hep-ph/0107308}}.

\bibitem{TaoHan1}
\hrefCMSnoop {} {T.~Han and B.~Zhang, ``{Signatures for Majorana neutrinos at
  hadron colliders}'',} \textit{ Phys. Rev. Lett.} \textbf{ 97} (2006) 171804,
  \href{http://dx.doi.org/10.1103/PhysRevLett.97.171804}{\doi{10.1103/PhysRevLett.97.171804}},
\href{http://www.arXiv.org/abs/hep-ph/0604064}{\texttt{ arXiv:hep-ph/0604064}}.

\bibitem{Aguila}
\hrefCMSnoop {} {F.~del Aguila, J.~A. Aguilar-Saavedra, and R.~Pittau, ``{Heavy
  neutrino signals at large hadron colliders}'',} \textit{ JHEP} \textbf{ 10}
  (2007) 047,
  \href{http://dx.doi.org/10.1088/1126-6708/2007/10/047}{\doi{10.1088/1126-6708/2007/10/047}},
\href{http://www.arXiv.org/abs/hep-ph/0703261}{\texttt{ arXiv:hep-ph/0703261}}.

\bibitem{TaoHan2}
A.~Atre\hrefCMSnoop {} { {et~al.}, ``{The Search for heavy Majorana
  neutrinos}'',} \textit{ JHEP} \textbf{ 05} (2009) 030,
  \href{http://dx.doi.org/10.1088/1126-6708/2009/05/030}{\doi{10.1088/1126-6708/2009/05/030}},
\href{http://www.arXiv.org/abs/0901.3589}{\texttt{ arXiv:0901.3589}}.

\bibitem{l3}
\hrefCMSnoop {} {{ L3} Collaboration, ``{Search for isosinglet neutral heavy
  leptons in Z$^0$ decays}'',} \textit{ Phys. Lett. B} \textbf{ 295} (1992)
  371,
\href{http://dx.doi.org/10.1016/0370-2693(92)91579-X}{\doi{10.1016/0370-2693(92)91579-X}}.

\bibitem{delphi}
\hrefCMSnoop {} {{ DELPHI} Collaboration, ``{Search for neutral heavy leptons
  produced in Z decays}'',} \textit{ Z. Phys. C} \textbf{ 74} (1997) 57,
\href{http://dx.doi.org/10.1007/s002880050370}{\doi{10.1007/s002880050370}}.

\bibitem{ATLAS}
\hrefCMSnoop {} {{ ATLAS} Collaboration, ``{Inclusive search for same-sign
  dilepton signatures in pp collisions at $\sqrt{s} = 7$ TeV with the ATLAS
  detector}'',} \textit{ JHEP} \textbf{ 10} (2011) 107,
  \href{http://dx.doi.org/10.1007/JHEP10(2011)107}{\doi{10.1007/JHEP10(2011)107}},
\href{http://www.arXiv.org/abs/1108.0366}{\texttt{ arXiv:1108.0366}}.

\bibitem{ATLAS_2}
\hrefCMSnoop {} {{ ATLAS} Collaboration, ``{Search for heavy neutrinos and
  right-handed W bosons in events with two leptons and jets in pp collisions at
  $\sqrt{s} = 7$ TeV with the ATLAS detector}'',} \textit{ Eur. Phys. J. C}
  \textbf{ 72} (2012) 2056,
  \href{http://dx.doi.org/10.1140/epjc/s10052-012-2056-4}{\doi{10.1140/epjc/s10052-012-2056-4}},
\href{http://www.arXiv.org/abs/1203.5420}{\texttt{ arXiv:1203.5420}}.

\bibitem{Wudka}
F.~del Aguila\hrefCMSnoop {} { {et~al.}, ``{Heavy Majorana neutrinos in the
  effective Lagrangian description: Application to hadron colliders}'',}
  \textit{ Phys. Lett. B} \textbf{ 670} (2009) 399,
  \href{http://dx.doi.org/10.1016/j.physletb.2008.11.031}{\doi{10.1016/j.physletb.2008.11.031}},
\href{http://www.arXiv.org/abs/0806.0876}{\texttt{ arXiv:0806.0876}}.

\bibitem{LRSM_1}
A.~Ferrari\hrefCMSnoop {} { {et~al.}, ``{Sensitivity study for new gauge bosons
  and right-handed Majorana neutrinos in $p p$ collisions at $\sqrt{s}$ = 14
  TeV}'',} \textit{ Phys. Rev. D} \textbf{ 62} (2000) 013001,
\href{http://dx.doi.org/10.1103/PhysRevD.62.013001}{\doi{10.1103/PhysRevD.62.013001}}.

\bibitem{LRSM_2}
K.~Huitu\hrefCMSnoop {} { {et~al.}, ``{Doubly charged Higgs at LHC}'',}
  \textit{ Nucl. Phys. B} \textbf{ 487} (1997) 27,
  \href{http://dx.doi.org/10.1016/S0550-3213(97)87466-4}{\doi{10.1016/S0550-3213(97)87466-4}},
\href{http://www.arXiv.org/abs/hep-ph/9606311}{\texttt{ arXiv:hep-ph/9606311}}.

\bibitem{beta_decay}
P.~Benes\hrefCMSnoop {} { {et~al.}, ``{Sterile neutrinos in neutrinoless double
  beta decay}'',} \textit{ Phys. Rev. D} \textbf{ 71} (2005) 077901,
  \href{http://dx.doi.org/10.1103/PhysRevD.71.077901}{\doi{10.1103/PhysRevD.71.077901}},
\href{http://www.arXiv.org/abs/hep-ph/0501295}{\texttt{ arXiv:hep-ph/0501295}}.

\bibitem{EW_limits}
\hrefCMSnoop {} {E.~Nardi, E.~Roulet, and D.~Tommasini, ``{New neutral gauge
  bosons and new heavy fermions in the light of the new LEP data}'',} \textit{
  Phys. Lett. B} \textbf{ 344} (1995) 225,
  \href{http://dx.doi.org/10.1016/0370-2693(95)91542-M}{\doi{10.1016/0370-2693(95)91542-M}},
\href{http://www.arXiv.org/abs/hep-ph/9409310}{\texttt{ arXiv:hep-ph/9409310}}.

\bibitem{ALPGEN}
M.~L. Mangano\hrefCMSnoop {} { {et~al.}, ``{ALPGEN, a generator for hard
  multiparton processes in hadronic collisions}'',} \textit{ JHEP} \textbf{ 07}
  (2003) 001,
  \href{http://dx.doi.org/10.1088/1126-6708/2003/07/001}{\doi{10.1088/1126-6708/2003/07/001}},
\href{http://www.arXiv.org/abs/hep-ph/0206293}{\texttt{ arXiv:hep-ph/0206293}}.

\bibitem{CTEQ5}
\hrefCMSnoop {} {{ CTEQ} Collaboration, ``{Global QCD analysis of parton
  structure of the nucleon: CTEQ5 parton distributions}'',} \textit{ Eur. Phys.
  J. C} \textbf{ 12} (2000) 375,
  \href{http://dx.doi.org/10.1007/s100529900196}{\doi{10.1007/s100529900196}},
\href{http://www.arXiv.org/abs/hep-ph/9903282}{\texttt{ arXiv:hep-ph/9903282}}.

\bibitem{pythia}
\hrefCMSnoop {} {T.~Sj{\"o}strand, S.~Mrenna, and P.~Z. Skands, ``{PYTHIA 6.4
  physics and manual}'',} \textit{ JHEP} \textbf{ 05} (2006) 026,
  \href{http://dx.doi.org/10.1088/1126-6708/2006/05/026}{\doi{10.1088/1126-6708/2006/05/026}},
\href{http://www.arXiv.org/abs/hep-ph/0603175}{\texttt{ arXiv:hep-ph/0603175}}.

\bibitem{Geant4}
\hrefCMSnoop {} {{ GEANT4} Collaboration, ``{GEANT4---a simulation toolkit}'',}
  \textit{ Nucl. Instrum. Meth. A} \textbf{ 506} (2003) 250,
\href{http://dx.doi.org/10.1016/S0168-9002(03)01368-8}{\doi{10.1016/S0168-9002(03)01368-8}}.

\bibitem{CMS_detector}
\hrefCMSnoop {} {{ CMS} Collaboration, ``{The CMS experiment at the CERN
  LHC}'',} \textit{ JINST} \textbf{ 3} (2008) S08004,
\href{http://dx.doi.org/10.1088/1748-0221/3/08/S08004}{\doi{10.1088/1748-0221/3/08/S08004}}.

\bibitem{pflow1}
\href {http://cdsweb.cern.ch/record/1194487} {{ CMS} Collaboration,
  ``Particle--Flow Event Reconstruction in {CMS} and Performance for Jets,
  Taus, and {\MET}'',} CMS Physics Analysis Summary CMS-PAS-PFT-09-001, (2009).

\bibitem{pflow2}
\href {http://cdsweb.cern.ch/record/1279341} {{ CMS} Collaboration,
  ``Commissioning of the Particle-Flow Reconstruction in Minimum-Bias and Jet
  Events from {\Pp\Pp} Collisions at 7 {TeV}'',} CMS Physics Analysis Summary
  CMS-PAS-PFT-10-002, (2010).

\bibitem{antiKT}
\hrefCMSnoop {} {M.~Cacciari, G.~P. Salam, and G.~Soyez, ``{The anti-$k_t$ jet
  clustering algorithm}'',} \textit{ JHEP} \textbf{ 04} (2008) 063,
  \href{http://dx.doi.org/10.1088/1126-6708/2008/04/063}{\doi{10.1088/1126-6708/2008/04/063}},
\href{http://www.arXiv.org/abs/0802.1189}{\texttt{ arXiv:0802.1189}}.

\bibitem{madgraph}
J.~Alwall\hrefCMSnoop {} { {et~al.}, ``{MadGraph 5: going beyond}'',} \textit{
  JHEP} \textbf{ 06} (2011) 128,
  \href{http://dx.doi.org/10.1007/JHEP06(2011)128}{\doi{10.1007/JHEP06(2011)128}},
\href{http://www.arXiv.org/abs/1106.0522}{\texttt{ arXiv:1106.0522}}.

\bibitem{lumi}
\href {http://cdsweb.cern.ch/record/1434360} {{ CMS} Collaboration, ``Absolute
  Calibration of the Luminosity Measurement at {CMS}: {W}inter 2012 Update'',}
  CMS Physics Analysis Summary CMS-PAS-SMP-12-008, (2012).

\bibitem{pdf}
M.~Botje\hrefCMSnoop {} { {et~al.}, ``{The PDF4LHC Working Group Interim
  Recommendations}'',} (2011).
\href{http://www.arXiv.org/abs/1101.0538}{\texttt{ arXiv:1101.0538}}.

\bibitem{jes}
\hrefCMSnoop {} {{ CMS} Collaboration, ``Determination of Jet Energy
  Calibration and Transverse Momentum Resolution in {CMS}'',} \textit{ J.
  Instrum.} \textbf{ 6} (2011) P11002,
  \href{http://dx.doi.org/10.1088/1748-0221/6/11/P11002}{\doi{10.1088/1748-0221/6/11/P11002}}.

\bibitem{diboson_xs}
\hrefCMSnoop {} {J.~M. Campbell, R.~K. Ellis, and C.~Williams, ``{Vector boson
  pair production at the LHC}'',} \textit{ JHEP} \textbf{ 07} (2011) 018,
  \href{http://dx.doi.org/10.1007/JHEP07(2011)018}{\doi{10.1007/JHEP07(2011)018}},
\href{http://www.arXiv.org/abs/1105.0020}{\texttt{ arXiv:1105.0020}}.

\bibitem{CLs1}
\hrefCMSnoop {} {A.~L. Read, ``{Presentation of search results: The $CL_s$
  technique}'',} \textit{ J. Phys. G} \textbf{ 28} (2002) 2693,
\href{http://dx.doi.org/10.1088/0954-3899/28/10/313}{\doi{10.1088/0954-3899/28/10/313}}.

\bibitem{CLs2}
\hrefCMSnoop {} {T.~Junk, ``{Confidence level computation for combining
  searches with small statistics}'',} \textit{ Nucl. Instrum. Meth. A} \textbf{
  434} (1999) 435,
  \href{http://dx.doi.org/10.1016/S0168-9002(99)00498-2}{\doi{10.1016/S0168-9002(99)00498-2}},
\href{http://www.arXiv.org/abs/hep-ex/9902006}{\texttt{ arXiv:hep-ex/9902006}}.

\bibitem{CLs3}
\href {https://cdsweb.cern.ch/record/1379837} {{ATLAS and CMS Collaborations},
  ``Procedure for the {LHC H}iggs boson search combination in Summer 2011'',}
  ATL-PHYS-PUB/CMS NOTE, {ATL-PHYS-PUB-2011011/CMS NOTE-2011/005}, (2011).

\end{thebibliography}\endgroup
